\documentstyle[preprint,eqsecnum,aps]{revtex}

\begin{document}
\draft
\preprint{TD Bank and UW}
\title{Generalizing Merton's approach of pricing risky debt:\\ some closed form results}
\author{D. F. Wang$^{1,2}$}
\address{$^1$Department of Statistics and Actuarial Science\\
University of Waterloo, 
Waterloo, ONT N2L 3E5 Canada
\\
$^2$Toronto Dominion Bank}
\date{1998}
\maketitle
\begin{abstract}
In this work, I generalize Merton's approach of pricing risky debt to the case
where the interest rate risk is modeled by the CIR term structure.
Closed form result for pricing the debt is given for the case where the
firm value has non-zero correlation with the interest rate.
This extends previous closed form pricing formular of zero-correlation case
to the generic
one of non-zero correlation between the firm value and the interest rate.  
\end{abstract}



One well known approach of pricing risky debt was pioneered by Merton a long time ago. 
Within the framework, one assumes a stochastic process for the firm
value and treats the 
risky debt as a option\cite{merton1,shimko,long,longstaff,zhou}. 
In Merton's original work\cite{merton1}, 
he assumed that the interest rate is constant
and that the default event of the debt can only occur at the time of
maturity. Closed form results for pricing the risky debt were given 
explicitly. As he has pointed out, one can also study the case
when stochastic interest rate is taken into account. This can easily
be achieved by using Merton's work that generalized Black-Scholes formular
of the option pricing with stochastic interest rate\cite{merton1}. 

However, one
has to make the assumption that the bond process of the stochastic
interest rate has a non-stochastic volatility which is
allowed to be deterministic time-dependent\cite{merton1}.
Therefore,  Shimko etc \cite{shimko} applied Merton's results to the case of
stochastic interest rate described by the Vasicek model, the bond process of which  
has a non-stochastic volatility. However, they were unable to handle the 
case where the interest rate is modeled with the CIR term structure, as 
the CIR interest rate model will give rise to a bond process having stochastic
volatility\cite{cox2}. 
It remained open whether one can give a similar closed form 
results for the risky debt when the interest rate risk is modeled with
CIR term structure. In one recent work, I gave the closed form pricing formular
for the risky debt in the case where CIR term structure is used for
the default-free interest rate. However, 
using the moment generating functional, one has
to assume that the firm value is uncorrelated with the interest rate\cite{wang}.
In this paper, I extend my closed form results of pricing 
risky debt to the generic
case where the correlation between the firm value and the interest rate is 
nonzero. 

Let us first assume a probability space denoted by
$(\Omega, P, \{F_t\}, F)$, with the filtration $\{F_t\}$.
Consider the value of the firm that is described by the following process
\begin{equation}
{dV\over V} =\mu dt + \sigma d Z_1, 
\end{equation}
where $Z_1$ is a Brownian motion in the probability space.
The interest rate process is assumed to be the one given by Cox-Ingersoll-Ross\cite{cox2}:
\begin{equation}
dr=(a -\beta r)dt  + \eta  dZ_2,
\end{equation}
where $\eta =\sigma_r   r^{1/2}$ with $\sigma_r$ as a constant. 
The co-quadratic variational process is $[Z_1,Z_2]=\rho t$. The correlation
coefficient $\rho$ is a non-zero constant during the following consideration.  

The assumptions in Merton's paper
are also made here\cite{merton1}. The firm value is assumed 
to be independent of 
its capital structure by assuming that MM theorem is valid.
 The firm issues debt and 
equity.  
The total value of the firm is 
the sum of equity and debt. The PDE satisfied by the equity 
is given by 
\begin{equation}
H_{\tau}={\sigma^2\over 2} V^2 H_{VV} +\rho \eta \sigma VH_{Vr}
+{\eta^2\over 2}H_{rr} +r VH_V +(\alpha -\beta r) H_r -r H
\end{equation}
where $H=H(V, r, T-t)$ and $\tau=T-t$ is time to the maturity, and 
$\alpha$ is sum of $a$ plus the constant representing 
the market price of the interest rate risk.
At $\tau=0$, the equity should satisfy the boundary condition
that $H=max(0, V(T)-B)$, where $B$ is the face value of the debt issued 
by the firm maturing  at time $T$. The risky debt price is therefore
given by $Y=V(t)-H(V,r,T-t)$. For simplicity, it is assumed here, 
as Merton did, 
that event of default of the risky debt can only occur at the time of 
maturity.   

Following the standard risk neutral approach, we write the equity 
price as below:
\begin{equation}
H=E^Q(e^{-\int_t^T r(s) ds} max(0,V(T)-B)|F_t),
\label{eq:risk1} 
\end{equation}
where the expectation $E^Q$ means that in the risk-neutral-adjusted world.
In this risk-neutral world, the firm value and the interest rate will follow
the stochastic differential equations as 
\begin{eqnarray}
&&d lnV=(r-{1\over 2} \sigma^2) dt +\sigma d\hat Z_1\nonumber\\
&& d r =(\alpha-\beta r )dt +\eta d\hat Z_2.
\label{eq:risk2}
\end{eqnarray}
Here, both $\hat Z_1$ and $\hat Z_2$ are Wiener processes in the risk-neutral
world, and the co-quadratic process is $[{\hat Z_1}, {\hat Z_2}]= \rho t $.
In principle, one can go on with Monte-Carlo simulation based on the above
formular Eq.(\ref{eq:risk1}) and Eq.(\ref{eq:risk2}).
However, we are interested in finding closed form result for
the equity price here. 

In the risk-neutral world, the two Brownian motions $\hat Z_1$ and
$\hat Z_2$ can be represented
in the following way. Suppose that $X$ and $Y$ are two independent Brownian
motions in the risk-neutral world. We can find such independent Brownian motions that  $\hat Z_1= \rho X + \sqrt{(1-\rho^2)}
Y$, and $\hat Z_2 = X$. The stochastic differential equation for the 
firm value is governed by
\begin{equation}
d lnV= [rdt + \sigma \rho dX -{1\over 2} \sigma^2\rho^2 dt]
-{1\over 2} \sigma^2 (1-\rho^2) dt +
\sqrt{(1-\rho^2)} \sigma dY,
\end{equation} 
where one is working in the risk-neutral world.

Conditional on that the sample path $\{ X\}$ of the Brownian motion $X$ is 
given for time interval $[t,T]$, let us consider the following  
expectation
\begin{equation}
h=E^Q(e^{-\int_t^T r(s) ds } max(V(T)-B,0) |F_t,\{X\})
=e^{-\int_t^T r(s) ds } E^Q(max(V(T)-B,0)|F_t, \{X\}) 
\end{equation}
This conditional expectation can be computed by standard way as dealing
with Black-Scholes case. It is found that 
\begin{equation}
h=e^{\sigma \rho \int_t^T dX}\cdot e^{-{1\over 2}\sigma^2\rho^2(T-t)}
 [ V(t)N(d_1)-Be^{-(\int_t^T r(s) ds-{1\over 2} \sigma^2\rho^2 (T-t)+ 
\sigma \rho \int_t^T dX)} N(d_2)], 
\end{equation}
where $N(d)$ is the standard accumulative function of 
normal distribution $N(0,1)$ 
and 
\begin{equation}
d_{1,2}= {ln(V(t)/B) + [\int_t^T r(s) ds -{1\over 2} \rho^2\sigma^2(T-t)
+ \sigma \rho \int_t^T dX 
\pm (1/2) \sigma^2 (1-\rho^2) (T-t)]\over \sigma\sqrt{(1-\rho^2) (T-t)}}. 
\end{equation} 
Let us denote $g(m,W)$ the probability density for $\int_t^T r(s) ds$
to be in the region $[m,m+dm]$, conditional on that $W=\int_t^T dX$ is fixed.
In the following we will show that $g(m,W)$ is independent 
of the variable $W$. 
The equity price of the firm can be represented as 
\begin{equation}
H=\int_{-\infty}^{+\infty} P(W) dW \int_0^{+\infty} g(m) dm [ e^{\sigma\rho W}
\cdot e^{-{1\over 2}\sigma^2\rho^2(T-t)} V(t) N(d_1) - Be^{-m} N(d_2) ], 
\label{eq:equity} 
\end{equation}
where $P(W)$ is normal distribution density, 
$P(W)= exp(-0.5 W^2/(T-t))/\sqrt{2\pi(T-t)}$, and 
\begin{equation}
d_{1,2}= {ln(V(t)/B) + [m -{1\over 2} \rho^2\sigma^2(T-t)+ \sigma \rho W 
\pm (1/2) \sigma^2 (1-\rho^2) (T-t)]\over \sigma\sqrt{(1-\rho^2) (T-t)}}.
\end{equation} 
The distribution function
$g(m)$ can be found from the moment generating functional.
For the Cox-Ingersoll-Ross term structure, the density function $g(m)$ can be
found easily. Consider the moment generating function
\begin{equation}
I(x)=E^Q(e^{-x \int_t^T r(s) ds} |F_t)=\int_0^\infty e^{-mx} g(m)dm,
\end{equation}
where $x$ is any non-zero numbers. Using the bond price of Cox-Ingersoll-Ross,
we find the moment generating function.
For CIR term structure, stochastic differential equation governing
the short rate is $dr = (\alpha-\beta r )dt +\sigma_r \sqrt{r} d\hat Z_2$,
will remain unchanged under the scaling transformation
\begin{eqnarray}
&&r\rightarrow x r\nonumber\\
&&\beta \rightarrow \beta\nonumber\\
&&\alpha\rightarrow x  \alpha\nonumber\\
&&\sigma_r\rightarrow x^{1/2} \sigma_r,
\end{eqnarray}
where $x$ is any positive real number.
Denote $D(r(t); \alpha, \beta, \sigma_r, T-t)=E^Q(e^{-\int_t^Tr(s)ds}|F_t)$
the riskless zero-coupon bond
price at time $t$ whose payoff at maturity $T$ is one. The exact closed
form of this zero-coupon bond price was provided explicitly\cite{cox2}.
We obtain
\begin{equation}
I(x) =D(xr(t), x\alpha, \beta, x^{1/2} \sigma_r, T-t)=
\int_0^\infty g(m) e^{-xm}dm.
\end{equation}
Doing inverse transformation, we will be able to find the density function
$g(m)$.

In order to see why $g(m,W)$ is independent of $W$, we first look at the 
discrete version of  $\int_t^T r(s) ds $. In the discrete version,
we see that $W$ dependence will only make higher order contribution.
Therefore, in the continuous limit, the density distribution of 
$\int_t^T r(s) ds $ for given $W$ will not have $W$ dependence. 
This is why we can represent the equity price in terms of two
double integral as Eq.(\ref{eq:equity}). The price of the risky debt
that pays $B$ dollars at time of maturity $T$ is simple given by
$V(t)-H$.

In summary, we have generalized Merton's approach of pricing risky debt
to the situation where the interest rate risk is modeled with the CIR
term structure. Exact closed forms for pricing risky debt are provided
explicitly. This goes beyond the situation where the riskless bond process
has non-stochastic volatility (such as for Vasicek interest rate model) 
and the option pricing can be handled by Merton's generalized Black-Scholes
method.   
 
Email address:d6wang@barrow.uwaterloo.ca.
I am indebted to Professors P. Boyle and D. McLeish for
the finance theories I learned from them. Conversations with
Prof. K. S. Tan of UW,
Dr. Hou-Ben Huang and Dr. Z. Jiang of TD Securities, Dr. Bart Sisk and Dr. A. Benn of TD Bank,
Dr. Daiwai Li
and Dr. Craig Liu of Royal Bank, Dr. ChongHui Liu  and Dr. J. Faridani
of Scotia Bank, are gratefully knowledged. I also wish to thank Dr. Rama Cont
for informative communication. The opinions of this article are those of the
author's, and they do not necessarily reflect the institutions the author is affiliated
with. Any errors of this article are mine.

After this work was completed, Dr.  E. Hofstetter of Imperial
College informed me of the most recent work of semi-reduced model approach of pricing
default debt\cite{cathcart}. I am grateful for the communication.


\begin{references}
\bibitem{merton1} R. C. Merton, ``On the pricing of corporate debt:
the risk structure of interest rates",
Journal of Finance, {\bf 29}, 449 (1974).
\bibitem{shimko} D. Shimko, N. Tejima and D. R. Van Deventer,
``The pricing of risk debt when interest rates are stochastic",
Journal of Fixed Income, 58-65, Sept. (1993).
\bibitem{long} F. Black and J. C. Cox, ``Valuing corporate securities:
some effects of bond indenture provisions",
Journal of Finance {\bf 31}, 351 (1976).
\bibitem{longstaff} F. Longstaff and E. Schwartz, ``A simple approach
to valuing risky fixed and floating rate debt",
working paper (UCLA), (1994).
\bibitem{zhou} C. Zhou, preprint of Board of Governers of Federal Reserve, (1998).
\bibitem{vasicek} O. Vasicek, ``An equilibrium characterization of the term
structure", Journal of Financial Economics {\bf 5}, 177 (1977).
\bibitem{cox2} J. C. Cox, J. E. Ingersoll and S. A. Ross,
``A theory of the term structure of interest rates".
Econometrica {\bf 53}, 385-407 (1985).
\bibitem{wang} D. F. Wang, preprint of TD Bank and Uiv. of Waterloo 
(to be published), (May 1998). 
\bibitem{cathcart} L. Cathcart et al, Journal of Fixed Income, Vol. 8 June 1998. 
\end{references}
\end{document}